# Probabilistic forecasts of sea ice trajectories in the Arctic: impact of uncertainties in surface wind and ice cohesion

**Sukun Cheng[1], Ali Aydoğdu[2], Pierre Rampal[1,3], Alberto Carrassi[4,5], and Laurent Bertino[1]**

[1] Nansen Environmental and Remote Sensing Center, Bergen, Norway; sukun.cheng@nersc.no; Laurent.Bertino@nersc.no
[2] Centro Euro-Mediterraneo sui Cambiamenti Climatici, CMCC, Bologna, Italy; ali.aydogdu@cmcc.it
[3] Univ. Grenoble Alpes/CNRS/IRD/G-INP, Institut de Géophysique de l'Environnement, Grenoble, France; pierre.rampal@univ-grenoble-alpes.fr
[4] Dept. of Meteorology and National Centre for Earth Observation, University of Reading, UK; n.a.carrassi@reading.ac.uk
[5] Mathematical Institute, University of Utrecht, The Netherlands; n.a.carrassi@uu.nl
* Correspondence: sukun.cheng@nersc.no

**Abstract:** We study the response of the Lagrangian sea ice model neXtSIM to the uncertainty in the sea surface wind and sea ice cohesion. The ice mechanics in neXtSIM is based on a brittle-like rheological framework. The study considers short-term ensemble forecasts of the Arctic sea ice from January to April 2008. Ensembles are generated by perturbing the wind inputs and ice cohesion field both separately and jointly. The resulting uncertainty in the probabilistic forecasts is evaluated statistically based on the analysis of Lagrangian sea ice trajectories as sampled by virtual drifters seeded in the model to cover the Arctic Ocean and using metrics borrowed from the search-and-rescue literature. The comparison among the different ensembles indicates that wind perturbations dominate the forecast uncertainty — i.e., the absolute spread of the ensemble —, while the inhomogeneities in the ice cohesion field significantly increase the degree of anisotropy in the spread – i.e., trajectories drift divergently in different directions. We suggest that in order to get enough uncertainties in a sea ice model with brittle-like rheologies, to predict sea ice drift and trajectories, one should consider using ensemble-based simulations where at least wind forcing and sea ice cohesion are perturbed.

## 1. Introduction

Sea ice covering the polar oceans is an important component of the Earth System. The dramatic changes of sea ice extent and volume in the Arctic have been regularly reported in the recent decades [1-3]. It is therefore crucial to understand the new state and characteristics of the Arctic sea ice and how it impacts the regional and global weather and climate [4]. Moreover, reliable sea ice forecasting systems are demanded for both operational and academic purposes [5]. For instance, a thinner sea ice cover offers opportunities to exploit trans-Arctic shipping routes but its faster dynamics also challenge the safety of operations [6].

One specific and important aim of sea ice models is to represent small-scale dynamics such as the formation of leads and ridges, together with the large-scale drift patterns of big ice plates and small ice floes. To achieve this goal, numerical models consider a momentum equation including a specific term involving the internal stresses that accounts for the rheology of the sea ice material. Several sea ice rheologies have been proposed for being used in continuum models; these include the Viscous-Plastic (VP, [7]), the Elastic-Plastic-Anisotropic (EPA, [8,9]), the Elasto-Brittle (EB [10]) and the Maxwell-Elasto-Brittle (MEB, [11]). In general, and particularly in fact of the various rheologies, evaluating and calibrating sea ice models against observations are necessary to improve their forecasting skill, as well as to identify and quantify the sources of uncertainty.

We study a recently raised sea ice model called neXt generation Sea Ice Model (neXtSIM) [12] that uses a novel approach to simulate the ice cover deformation, damage, and healing processes. neXtSIM is a full dynamical-thermodynamical Lagrangian sea ice model that has been developed with the aim of better simulating sea ice dynamics, and sea ice trajectories in particular. A former version of neXtSIM included the EB rheology and, as demonstrated by Rampal et al. [13] and [14] using Synthetic Aperture Radar (SAR) data, neXtSIM accurately simulates the large-scale sea ice drift and Lagrangian diffusion properties, the sea ice cover thickness and extent, as well as the spatial scaling of the sea ice deformation. The rheology implemented in neXtSIM has been recently updated from EB to MEB, hereafter respectively denoted as neXtSIM-EB and neXtSIM-MEB. The MEB provides extra viscous mechanisms in the stress-strain relation, which reduce to the EB if the viscous relaxation time scale of the stresses for undamaged sea ice is taken sufficiently large. neXtSIM-MEB shows remarkable capabilities at reproducing the observed characteristics of sea ice kinematics and dynamics [15], in particular the space-time coupling of sea ice deformation scaling invariances [12] — a property never reproduced by a sea ice model before.

Sea ice drift forecasts are affected by multiple sources of uncertainties. The surface winds are one of the most important external forces driving the motion of the sea ice in the central Arctic [16]. At the same time, the uncertainties in the atmospheric reanalysis in the Arctic, from where such wind forcings are taken, are higher than that at the mid-latitudes, and observations are insufficient to estimate the statistical characteristics (scale, amplitudes) of the errors. Rabatel et al. [17] investigated the sensitivity of sea ice drift using neXtSIM-EB to the uncertainties of the surface winds. They concluded that in regions of highly compact ice cover the accuracy of surface wind forcing and sea ice rheology are both important for the probabilistic forecast skill of sea ice trajectories.

Observations of ice conditions are also essential to initialize sea ice simulations. Currently, the resolution of ice concentration data is too coarse to provide structures smaller than the very broad leads [18]. Even if they were available at very high resolution, other linear kinematic features (LKFs) like cracks and ridges resolved in the model would still remain unseen and uncertain.

Modelling the damage of the sea-ice brings another source of uncertainty, that is also particularly difficult to constraint given that the damage is not observable. Although desirable, directly perturbing the LKFs in the damage variable would need a very different approach, using random objects rather than random variables. The brittle class of rheological frameworks (e.g., EB and MEB) use the Mohr-Coulomb failure criterion to weaken the mechanical strength of sea ice by increasing the local ice damage level (see e.g., [12]); this criterior is regulated in the model by the cohesion parameter. Sea ice cohesion and friction coefficient in the failure criterion are important factors affecting in sea ice modelling. Sea ice cohesion sets the local resistance of sea ice to pure shear deformation until break up. The optimal value of the cohesion to be used in models should depend on the spatial scale and local inhomogeneities in the ice microstructure [19]. Nevertheless, those are not explicitly represented, and the cohesion is therefore commonly assumed constant in time and homogeneous in space. The limitations and inadequacy of this choice was put forward by Bouillon and Rampal [20] that showed how the cohesion in neXtSIM significantly affects the properties of the sea ice deformation patterns both in time and space, suggesting it may also impact the sea ice drift. The second key parameter driving sea damage, the friction coefficient, was set constant (= 0.7) in [20], and as such independent of the angle of friction. This value, that we also use in this work, is consistent with laboratory measurements (e.g., [21]) and suggested being scale-independent [22]. In a pragmatic perspective, given that not all sources of uncertainty are accessible as inputs of the model, only a few of them should be considered for ease of interpretation. We thus intentionally focus on the uncertainties from winds and sea ice cohesion as their importance in the short time scales sea ice modelling has been put demonstrated in previous works ([17,20]).

The present study extends the work of [17] along two directions. First, it studies the effect of the MEB rheology compared to the EB used in [17]. Second, we assess and compare the response of neXtSIM-MEB to uncertainties in the surface wind forcing and in the sea ice cohesion. We also study the predictive skill in terms of the forecast of sea ice trajectories. The paper is organized as follows. Section 2 presents the methodology, based on the analysis of the ensembles of virtual drifter trajectories. Section 3 describes the experiment setup and how we generated the different sets of ensembles used in this study. Section 4 presents the results of the ensemble forecasts while a discussion is given in Section 5. Section 6 gives the conclusions and perspectives for the ensemble forecasting and data assimilation.

## 2. Methodology

The Lagrangian trajectory simulations adhere to [17]: ensemble forecasts are conducted with multiple virtual drifters that are seeded in the model at the location and time of interest. The successive positions of the virtual drifters are updated at each time step according to the simulated velocity field, leading to a set of Lagrangian trajectories; the model response to uncertainty sources is estimated statistically using these trajectories. We denote $\boldsymbol{g}_i(x_0, t_0, t)$ as the drifter position of the $i$-th ensemble member at time $t$, with $\boldsymbol{x}_0$ being the initial position where the drifter is deployed at initial time $t_0$, and $\boldsymbol{B}(t)$ as the barycenter — the mean of the $N$ ensemble positions:

$$\boldsymbol{B}(t) = \frac{1}{N}\sum_{i=1}^{N} \boldsymbol{g}_i(\boldsymbol{x}_0, t_0, t). \quad (1)$$

The distance between the drifter positions in the $i$-th ensemble member and the barycenter $\boldsymbol{B}(t)$ is defined as

$$b_i(t) = \|\boldsymbol{g}_i(\boldsymbol{x}_0, t_0, t) - \boldsymbol{B}(t)\|, (i = 1, 2, \dots, N), \quad (2)$$

where $\|\cdot\|$ is the L-2 norm operator. The ensemble spread $\sigma_b$ is calculated as the standard deviation of $b_i(t)$. We define the position error, $\boldsymbol{e}(t)$, as the difference between $\boldsymbol{B}(t)$ and a reference drifter position $\boldsymbol{O}(t)$ at time $t$ as

$$\boldsymbol{e}(t) = \boldsymbol{B}(t) - \boldsymbol{O}(t), \quad (3)$$

where $\boldsymbol{O}(t)$ could be either from an observation or a reference numerical simulation, and $\|\boldsymbol{e}(t)\|$ indicates the distance between the ensemble mean and the reference drifter position. We define the parallel, $e_{\parallel}(t)$, and perpendicular, $e_{\perp}(t)$, components of $\boldsymbol{e}(t)$ onto the orthonormal basis centered on $\boldsymbol{O}(t)$, where the parallel direction is taken from $\boldsymbol{x}_0$ to $\boldsymbol{B}(t)$. The choice of the coordinate system follows [17] for comparing the results in Section 4.1. The parallel and perpendicular components describe the advection and diffusion of the virtual drifters in ensemble predictions.

The forecast uncertainty in [17] is described by an anisotropic search ellipse, defined by the smallest ellipse encompassing all ensemble drifter positions, centered at $\boldsymbol{B}(t)$ and with its long axis pointed to the origin $\boldsymbol{x}_0$. In this study, we rather define anisotropy in the way that follows the possible rotations of the ensemble of virtual drifters. We construct an anisotropic search ellipse that fits the ensemble of drifter positions using a bivariate Gaussian Mixture distribution. The resulting cumulative density function gives the center of the ellipse, and the ellipse size is determined to include 99% of the probability density. Our definition implies that ellipses are smaller than those in [17] and are equal whenever the segment between $\boldsymbol{x}_0$ and $\boldsymbol{B}(t)$ overlaps with the long axis. The area and the aspect ratio of the ellipse are obtained as

$$A = \pi ab \text{ and } R = a/b, \quad (4)$$

where $a$ and $b$ are the semi-major and semi-minor axes of the ellipse, respectively. The area indicates an absolute measure of the uncertainty and the aspect ratio indicates the anisotropy of uncertainty, which are used to assess the model sensitivity in section 4.2.

For the sake of clarity, we denote the spatial average over all buoys of a general quantity $f$ at time $t$ as

$$\langle f(t)\rangle_S = \frac{1}{M}\sum_{i=1}^{M} f(\boldsymbol{x}_i, t), \quad (5)$$

where $M$ is the total number of buoys, and $\boldsymbol{x}_i$ is the position of the $i$-th buoy. Similarly, we denote the average of $f$ over all the periods $N_t$ at position $\boldsymbol{x}$ as

$$\langle f(x)\rangle_T = \frac{1}{N_t}\sum_{j=1}^{N_t} f(x, t_j) \quad (6)$$

to even out the fluctuations of weekly environmental conditions, where subscript $j$ indicates the $j$-th period.

## 3. Experiment setup

### *3.1. Ocean and atmosphere forcing fields*

The neXtSIM model is used in a stand-alone configuration, driven by ocean and atmospheric reanalysis products. In this study, the ocean forcing comes from TOPAZ4, the latest version of a coupled ocean-sea ice data assimilation system covering the North Atlantic and Arctic Oceans [23]. TOPAZ4 is based on the Hybrid Coordinate Ocean Model (HYCOM) and assimilates both ocean and sea ice observations using the ensemble Kalman filter (EnKF) [24]. The ocean forcing provided by TOPAZ4 includes the sea surface height, the current velocity at 30 m depth, the sea surface temperature and salinity, all given as daily mean values in average horizontal resolution of 12.5 km [13]. The atmospheric forcing comes from the Arctic System Reanalysis (ASR, [25]) at a horizontal resolution of 30 km with 3 hours frequency. The variables used to force the model are 10 m wind velocity, 2 m temperature and mixing ratio, the mean sea level pressure, total precipitation and the fraction of that which is snow, and the incoming short-wave and long-wave radiations (see also [26]).

### *3.2. Simulation setup*

We focus on simulating sea ice drift in the Pan-Arctic Ocean during winter, when the extent and the volume of ice are close to their annual maximum (cf. the distribution of averaged ice thickness and concentration from 1 January to 28 April 2008 in Figure 1(a)(b), respectively). This provides abundant data of compact sea ice, enhancing the impact of ice cohesion on the sea ice drift.

The ensemble simulations are initialized from a deterministic neXtSIM simulation which is spun up from 1 January 2007 to 28 April 2008. The initial sea ice concentration and thickness on 1 January 2007 are taken from the TOPAZ4 reanalysis while the sea ice drift and damage are set to zero everywhere. We conduct a series of ensemble simulations from 1 January to 28 April 2008 restarted every 9 days from the deterministic simulation and lasting for 10 days. We consider that the sea ice drift is mainly influenced by the sea ice dynamics, while the effect of thermodynamics is negligible in such a short time scale. The key configurations in the simulations are summarized in Table 1.

neXtSIM is spatially discretized using the finite element method on a triangular mesh which is adaptive in time, meaning that it is automatically regenerated when and where triangles are too distorted using an efficient remeshing algorithm locally [13]. The nominal mesh resolution (defined as the mean of the square root of triangles surface areas) used for the experiments is about 7.5 km. The forcing fields are interpolated onto the center of the triangular elements during the model integration. neXtSIM is recently enhanced in computational efficiency, thanks to the parallelization of the code [27]. The simulations are carried out in parallel and each 10-day forecast takes ~0.5-hour wall clock time / ~16 CPU hours on 32 processors of a Lenovo supercomputer. Note that the random model uncertainties accumulate and grow over the duration of the ensemble forecast, like the forecast errors accumulate during an operational forecast.

### *3.3. Lagrangian trajectories*

We seed three sets of virtual drifters in the model at the initial time of each run. The first set of drifters are initially seeded at the same locations as the International Arctic Buoy Program (IABP)[1] buoys. It is used to evaluate the skill of the model simulation of observed sea ice trajectories (section 4.1). In the second set, the initial positions of the drifters are regularly spaced by 50 km and covering the whole Arctic Ocean. This set is used to calculate model ensemble statistics (section 4.2). The third set, similar to the second except that drifters' initial locations, are matching with the Ocean and Sea Ice Satellite Application Facility (OSI SAF) ice drift product[2] (which provides estimated drift vectors separated by 62.5 km, [28]). Drifters of this set are redeployed every second day throughout the

1 http://iabp.apl.washington.edu

2 http://www.osi-saf.org/?q=content/global-low-resolution-sea-ice-drift

simulation to be compared consistently with the OSI SAF observations. This last set is used to calibrate some of the model parameters (section 3.4).

Figure 1(c) shows the trajectories of the IABP buoys from 1 January to 28 April 2008 (red tracks) and the initial layout of the regularly spaced virtual drifters separated by 50 km (green dots). These virtual drifters are located further than 100 km away from the nearest coast as the same experimental choice was done by, e.g., [17], in order to prevent the potential bias in the statistics due to the extremely anisotropic drifts present in the vicinity of coastlines. For the same reason, we only account for IABP buoys located 100 km away from the coast in the following analyses.

In Figures 1(a)(b), we show general sea ice conditions (concentration and thickness) from TOPAZ4. The IABP buoys are deployed in pack ice areas where sea ice concentration is higher than 95%. The averaged ice thickness grows from 1.7 m in January to a plateau around 2.2 m since March, following the normal seasonal cycle of ice thickness. The winds directions are shifting and a particular episode of strong winds towards Greenland occurs in mid-March as we will show in Figure 4.

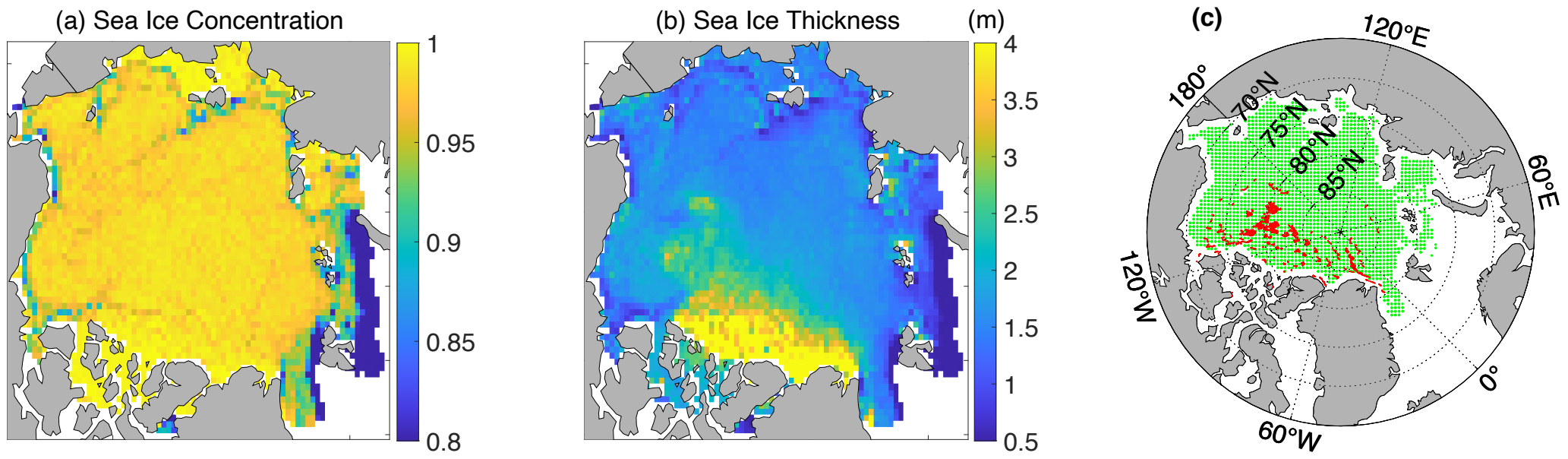

**Figure 1**. Distributions of sea ice concentration (a) and thickness (b) from the deterministic run averaged from 1 January to 28 April 2008. (c) Illustration of initial positions of regularly spaced virtual drifters (green dots) and the IABP buoy trajectories overlaid from 1 January to 28 April 2008 (red dots).

### 3.4. *Air drag coefficient optimization*

When using a forced —stand-alone— sea ice modeling system as we do here, the momentum fluxes from the wind and ocean currents to the sea ice need to be calibrated. This can be achieved by tuning the value of the air and/or water drag coefficients denoted $c_a$ and $c_w$, respectively (see Table 2). Following the calibration method presented in [13], the air drag coefficient is optimized by comparing the simulated sea ice drift with observations from the OSI SAF dataset [29], but only where sea ice is in "free drift" while keeping the default value of the ocean drag coefficient unchanged. We identify the *free-drift* events when and where the simulated sea ice drift obtained, when the sea ice rheology is activated, differs by less than 10% of that obtained when it is deactivated. We use an additional criterion on the sea ice speed, which must be sufficiently large, i.e., between 7 and 19 km/day as used in [13]. The selected drift velocity vectors are compared against the OSI SAF drift vectors and for the two components U and V separately in the horizontal and vertical directions defined in a polar stereographic projection of the Arctic.

We calculate the correlation coefficient and the Root Mean Square Error (RMSE) for each component. The results obtained from a set of simulations using values of air drag coefficient ranging from 0.003 to 0.008 are presented in Figure 2. We find an optimal value of $c_a$ about 0.0055, which is in between previously optimized values of 0.0051 and 0.0076 reported by [13] and [17]. We note that these latter studies were using the same atmospheric forcing dataset but the neXtSIM-EB model instead of the neXtSIM-MEB.

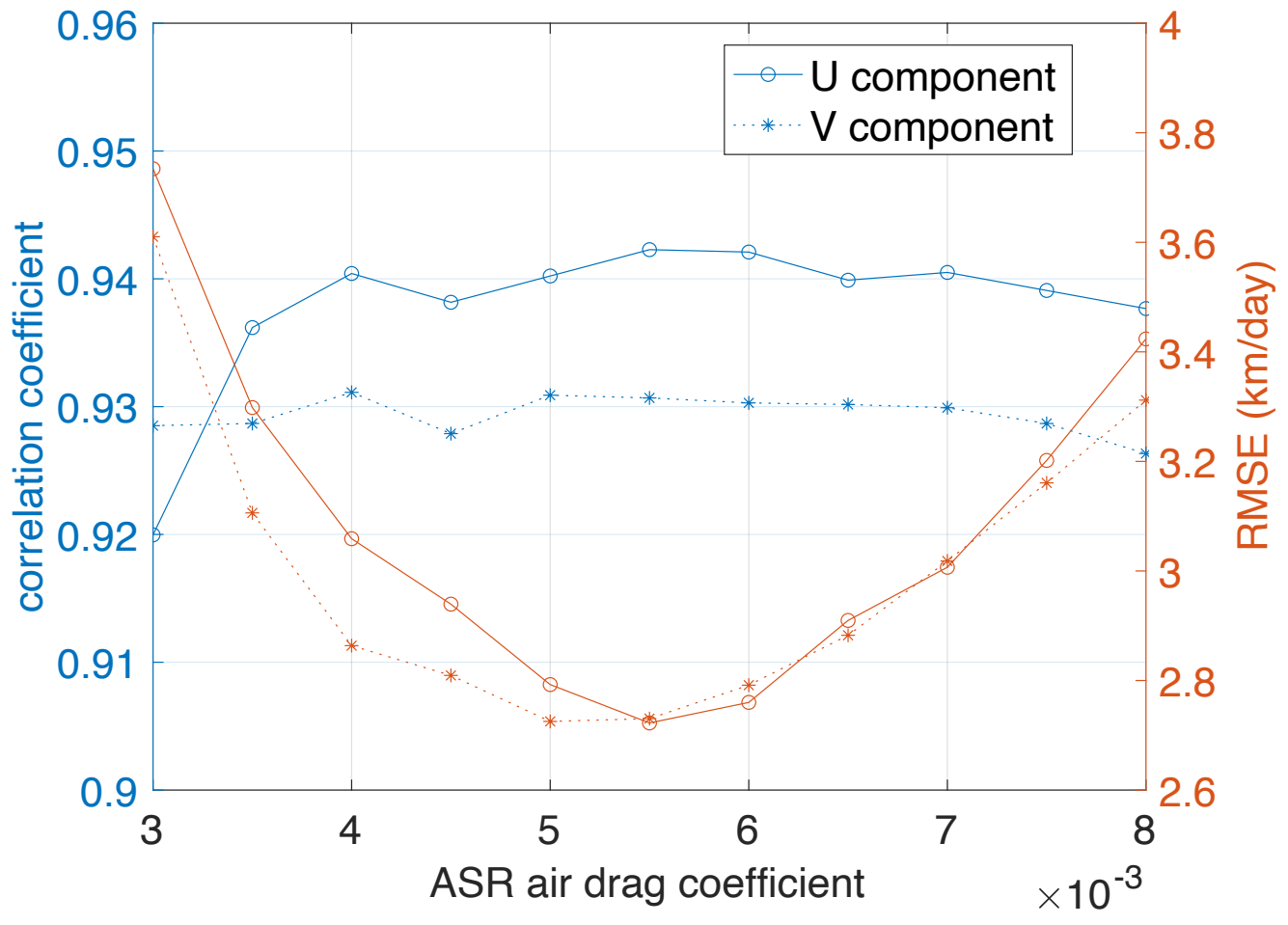

**Figure 2**. Correlation coefficients (left axis) and RMSE (right axis) calculated when comparing the simulated and observed (OSI SAF) ice drift velocity vectors when using different air drag coefficient values (horizontal axis).

### *3.5. Ensemble experiments by perturbing wind/cohesion sources*

The impact of the two sources of uncertainty under consideration is studied by conducting ensembles of simulations where the wind forcing and/or the cohesion are perturbed. Except for these two, the model configuration for each ensemble member is the same. Note that the nature of the two uncertainties is different since the former is an external forcing while the latter is a parameter of the sea ice model. However, uncertainties in either of them contributes to the forecast errors, and we aim here at quantifying their individual and joint effects.

The wind-perturbation process adopted in this study is identical to [17], inherited from the TOPAZ4 data assimilation system [23]. The perturbations are non-divergent time-correlated wind fields with a decorrelation time scale of 2 days and a horizontal decorrelation length scale of 250 km. A non-divergence constraint is imposed for the perturbation so that the perturbed winds are no more divergent than the original winds: an increase in wind divergence would inevitably result in breaking the sea ice cover and lead to a large ensemble bias. The accumulation of these random perturbations along all model simulations causes the ensemble members to diverge from each other. We adjust the wind speed variance of the perturbation to 3 $m^2/s^2$ to generate sufficient ensemble spread. The perturbation process is applied online to every input wind field.

The cohesion being an intrinsic parameter of the sea ice model, its values are perturbed only once at the initial time and then kept constant throughout the ensemble forecast. We first conduct a broad screening using different cohesion values (5.5, 11, 16.5, 22, 27.5, 33, 38.5, 44, 49.5 and 55 kPa), keeping the model configuration almost identical to the deterministic run except for the optimal air drag coefficient $c_a$= 0.0055 determined above. We note that the cohesion is a scale dependent property of sea ice, and these values are scaled at 7.5 km mesh resolution from laboratory value using the same relationship as in [12]. The cohesion optimization benefits from the IABP buoys because they are mostly in the thick ice regions, where the cohesion plays a major role. Figure 3 shows that the spatially averaged errors $\langle\|\boldsymbol{e}(t)\|\rangle_S$, $\langle e_{\parallel}(t)\rangle_S$ and $\langle e_{\perp}(t)\rangle_S$ are the lowest when setting the ice cohesion between 20 and 40 kPa, which is consistent with the default — but not robustly optimized — cohesion of 25 kPa used in [12] for their neXtSIM-MEB simulations.

To further investigate the effect of sea ice cohesion uncertainties on simulated sea ice trajectories, we conduct ensemble simulations using inhomogeneous initial cohesion fields, so-called perturbed cohesion fields hereafter. At the start of each ensemble simulation, the cohesion value is chosen randomly between 20 and 40 kPa in each element of the model mesh following a uniform distribution and without spatial correlation. Afterwards, the values are kept constant during the simulation except in case of remeshing: the cohesion values in the new elements created by the remeshing process are calculated as the mean of the cohesion of their nearest neighbor elements.

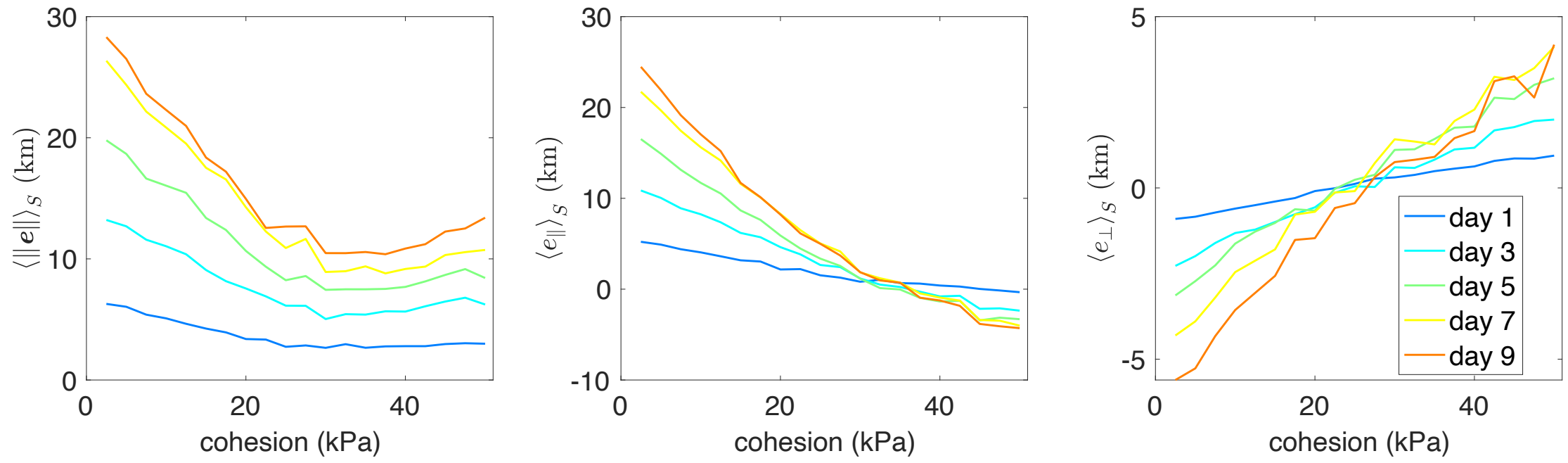

**Figure 3.** Bias between observed trajectories of the IABP buoys and the corresponding simulated drifter trajectories obtained from a set of simulations with different initial — homogeneous — cohesion fields. The spatial averaged forecast errors $\langle\|\boldsymbol{e}(t)\|\rangle_S$, $\langle e_\parallel(t)\rangle_S$ and $\langle e_\perp(t)\rangle_S$ indicate the averaged errors over all IABP buoys defined in Section 2. The colors correspond to different lead times at which the biases are calculated.

Three experiments are conducted by applying either the wind perturbations alone (WIND), the cohesion perturbations alone (COHESION), or joint wind and cohesion perturbations (JOINT) to generate the ensembles. Each ensemble contains 20 members, and each member contains 10-day forecasts for the 13 successive periods. Note that in the Monte Carlo techniques, the estimates of ensemble prediction converge by increasing the ensemble size. In our study, the convergence is achieved when the ensemble size exceeds 20 (not shown). The experiment setup is recapped in Table 1. Values of the model parameters used for running the forecasts are given in Table 2.

**Table 1.** Experiment setup.

| Ensemble acronym | WIND | COHESION | JOINT |
|---|---|---|---|
| Ensemble generation | Perturbation of winds using random wind fields | Perturbation of ice cohesion initialized from a random uniform distribution | Joint winds perturbation and ice cohesion perturbation |
| Ensemble size | 20 | 20 | 20 |
| Initial dates (DD-MM-2008) of each 10-day forecast | 01-01, 10-01, 19-01, 28-01, 06-02, 15-02, 24-02, 04-03, 13-03, 22-03, 31-03, 09-04, 18-04 | | |
| Atmospheric forcing | ASR reanalysis | | |
| Oceanic forcing | TOPAZ4 reanalysis | | |
| Initial sea ice thickness and concentration | | | |

**Table 2.** List of parameters in the model and their values for the simulations used in this study.

| Symbol | Name | Value | Unit |
|---|---|---|---|
| $\rho_a$ | Air density | 1.3 | kg/m³ |
| $c_a$ | Air drag coefficient | 0.0055 | - |
| $\theta_a$ | Air turning angle | 0 | degree |
| $\rho_w$ | Water density | 1025 | kg/m³ |
| $c_w$ | Water drag coefficient | 0.0055 | - |
| $\theta_w$ | Water turning angle | 25 | degree |
| $\rho_i$ | Ice density | 917 | kg/m³ |
| $\rho_s$ | Snow density | 330 | kg/m³ |
| $\nu$ | Poisson ratio | 0.3 | - |
| $\mu$ | Internal friction coefficient | 0.7 | - |
| $Y$ | Elastic modulus | 596 | MPa |
| $\Delta x$ | Nominal mesh resolution | 7.5 | km |
| $\Delta t$ | Time step | 200 | s |
| $T_d$ | Characteristic time for damaging | 20 | s |
| $\lambda_0$ | Undamaged relaxation time | $10^7$ | s |
| $\alpha$ | Compactness parameter | -20 | - |

## 4. Results

In this section, we compare the impacts of the perturbation methods. In Section 4.1, results of the experiments are presented by comparing the ensemble spread and bias to the IABP buoys. Section 4.2 presents the analysis of the ensemble statistics of the virtual drifters' trajectories over the whole Arctic region. Note that neXtSIM only tracks the positions of virtual drifters when the concentration within a model grid cell is greater than 15%. In the following analysis, we calculate the statistics of the simulated ensembles only from the drifter trajectories spanning the full forecast period of 10 days.

### *4.1. Comparison of the simulated sea ice trajectories against the IABP dataset*

The spatial averaged forecast errors over the IABP buoys are presented in Figure 4(a). The top panel shows the mean and standard deviation of the spatially averaged errors $\langle\|\boldsymbol{e}(t)\|\rangle_S$ as the solid lines and error bars respectively, marked by different colors, one for each experiment. The behavior of the three ensembles relative to each other is generally in agreement across successive forecasts: the horizontal variability range of $\langle\|\boldsymbol{e}(t)\|\rangle_S$ in COHESION is generally the smallest, contained within that of WIND and/or JOINT that are similar to each other. This implies the errors are generally larger when surface winds are perturbed and more variable geographically. One exception occurs in the last period during 18 – 27 April when COHESION gives the largest errors $\langle\|\boldsymbol{e}(t)\|\rangle_S$. It could be related to the seasonal change of Arctic sea ice. The period during 13 – 22 March also stands out as the difference between WIND and JOINT that $\langle\|\boldsymbol{e}(t)\|\rangle_S$ is larger than usual. We speculate that it is due to a wind event during 11 – 18 March, when wind is strong and in statistically uniform direction around the IABP buoys' deployment area shown in Figure 4(b), which gives the daily-averaged horizontal wind velocities from the ASR wind dataset over the IABP buoys trajectories. This point is further discussed in Section 5.

The parallel and perpendicular error components, $\langle e_\parallel(t)\rangle_S$ and $\langle e_\perp(t)\rangle_S$, are shown in the middle and bottom panels of Figure 4(a) respectively, where only the means are given as the solid lines. As defined in Section 2, a positive/negative $e_\parallel(t)$ indicates that the virtual drifters move faster/slower than the real buoy along the direction from $x_0$ to $B(t)$. And a positive/negative $e_\perp(t)$ indicates that the virtual drifters move to the right/left side of a vector from $x_0$ to $B(t)$. Again, the errors from the three ensembles generally follow the same order across successive forecast runs. It also shows that $\langle\|\boldsymbol{e}(t)\|\rangle_S$ has its largest contribution from its parallel component $\langle e_\parallel(t)\rangle_S$, even more so when the trajectories are erratic.

The quantities in Figure 4(a) are further averaged over the 13 time periods as $\langle\langle\|\boldsymbol{e}(t)\|\rangle_S\rangle_T$, $\left\langle\langle e_\parallel(t)\rangle_S\right\rangle_T$ and $\langle\langle e_\perp(t)\rangle_S\rangle_T$ shown in Figure 4(c). The results from WIND and JOINT are in general agreement with Figure 15 of [17]: the forecast tends to drift too fast and to the right of the observed trajectories on average, although the time series in Figure 4(a) shows the errors differ significantly from one forecast to another and are flow dependent. The WIND and JOINT ensemble tend to show longer drifts while the COHESION ensemble drifts slightly more to the right of the trajectory. Wind perturbations are constructed to have zero-mean components in the x- and y-directions of the grid, however since the absolute velocity is not a linear function of the x- and y-components, the ensemble average velocity of the perturbed winds is larger than the original

unperturbed winds. This may explain the longer drifts in the experiments using wind perturbations. The evolution of the perpendicular error (to the right, then to the center regarding observations) does not appear like a recurrent feature in Figure 4(a) and is probably not significant.

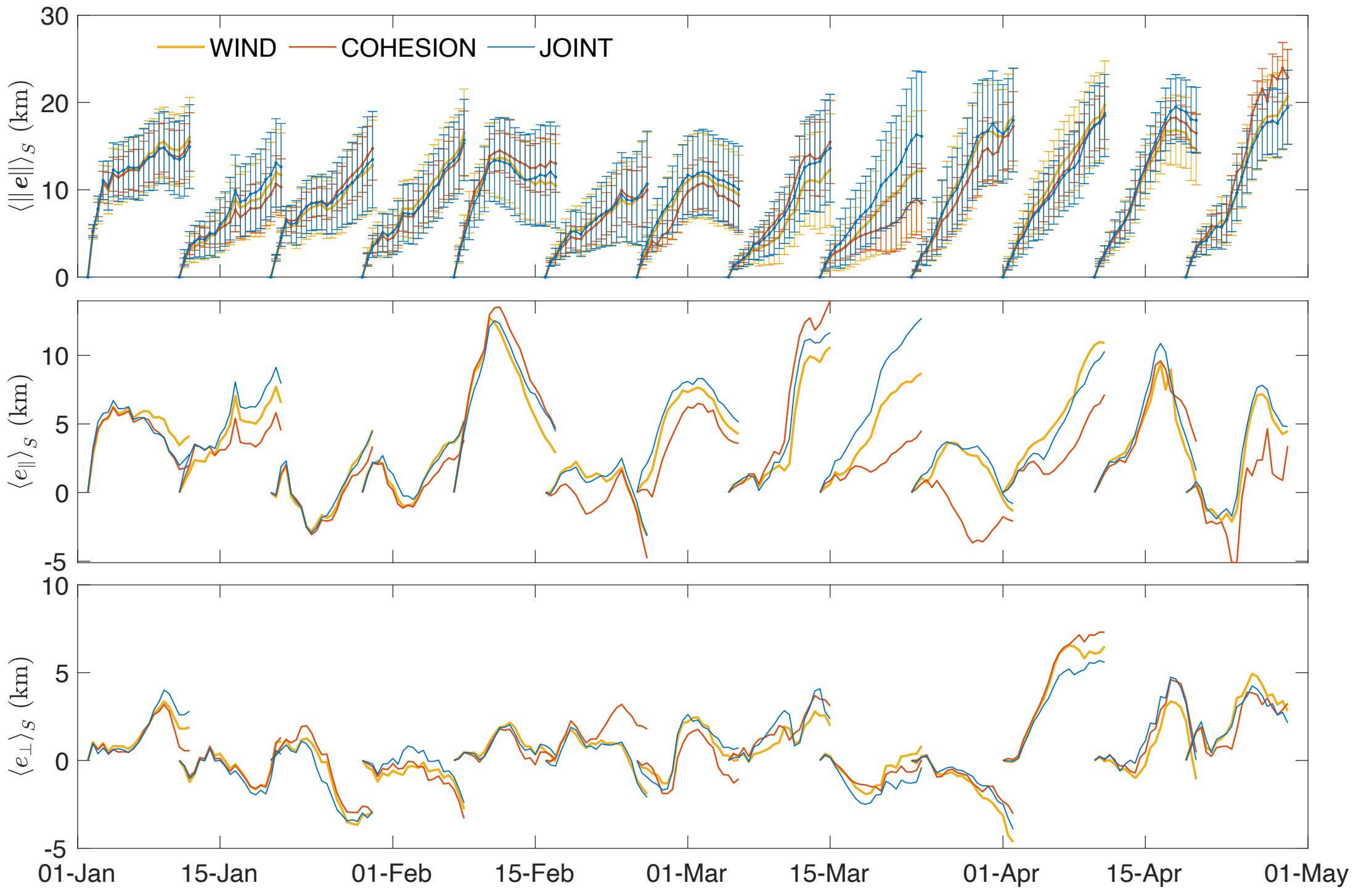

(a) Spatially averaged errors $\langle\|\boldsymbol{e}(t)\|\rangle_S$, and their parallel and perpendicular components $\langle e_\parallel(t)\rangle_S$ and $\langle e_\perp(t)\rangle_S$. Error bars are omitted from the latter two to enhance clarity.

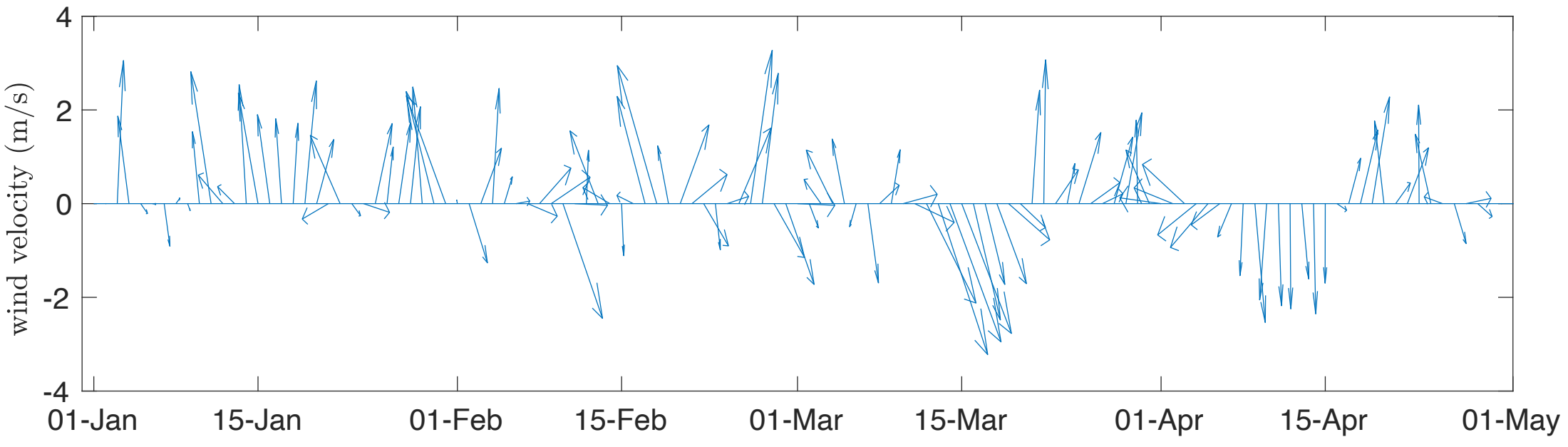

(b) Feather plot of daily averaged wind velocities over the IABP buoys positions.

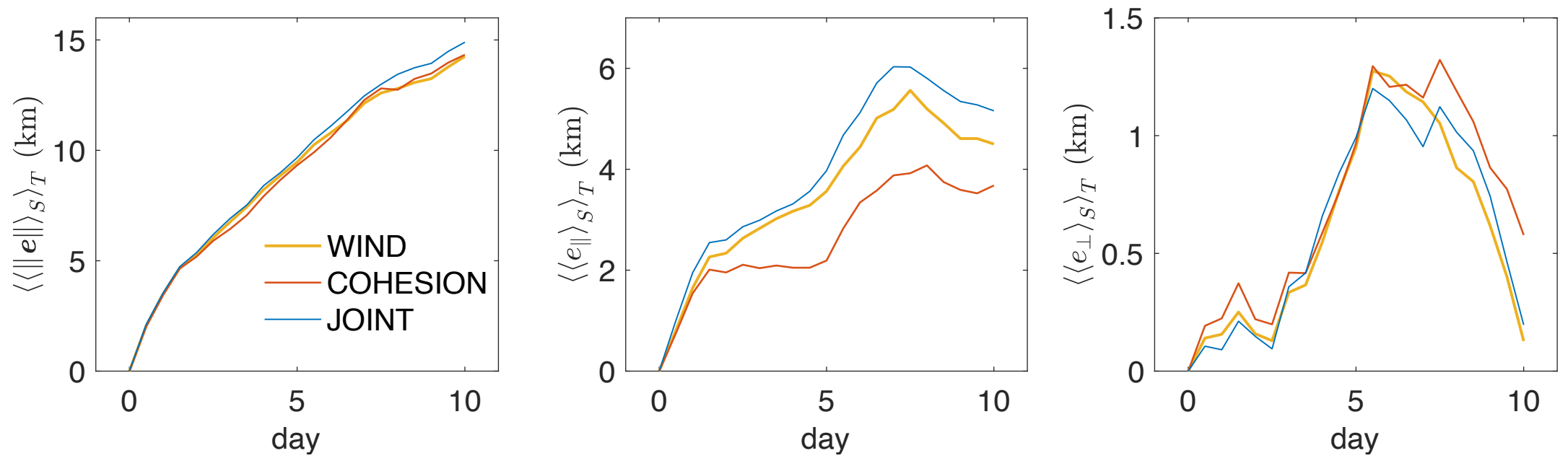

(c) Averages of errors in panel (a) over the 13 time periods.

**Figure 4.** Errors of positions between the ensemble mean of virtual drifters and the drifting buoys as a function of time.

A complementary indicator of the quality of the ensemble forecasts is the ratio of ensemble spread over the root-mean-square error (RMSE) between the ensemble mean forecast and the observation, i.e., $\langle \sigma_b(t) \rangle_S / \langle \|\boldsymbol{e}(t)\| \rangle_S$ taken as a spatial average. A ratio less than 1 indicates an under estimation of errors in the ensemble forecasts compared to the actual position errors. As shown in Figure 5(a), the ratio in our case is most often below 0.5 but shows strong temporal variations, often in parallel between the three runs, and increases briefly in mid-March. The low ratio indicates that the ensemble spread is insufficient to encompass the real buoys, a result that agrees with [12]. The ratio could be increased by using more efficient ways to perturb the surface winds and cohesion studied in this work, as well as introducing other sources of uncertainty, as mentioned in the introduction. The spatially averaged spread $\langle \sigma_b(t) \rangle_S$ is given in Figure 5(b). The spread of WIND and JOINT are almost identical and larger than that of COHESION, excepted for the second week of March. We will come back to this in section 4.2. The phenomena of rapid increase of the spread in 11 – 14 March in Figure 5(b) and a peak of $\langle \sigma_b(t) \rangle_S / \langle \|\boldsymbol{e}(t)\| \rangle_S$ in WIND and JOINT on 15 March in Figure 5(a) are further discussed in Section 5.

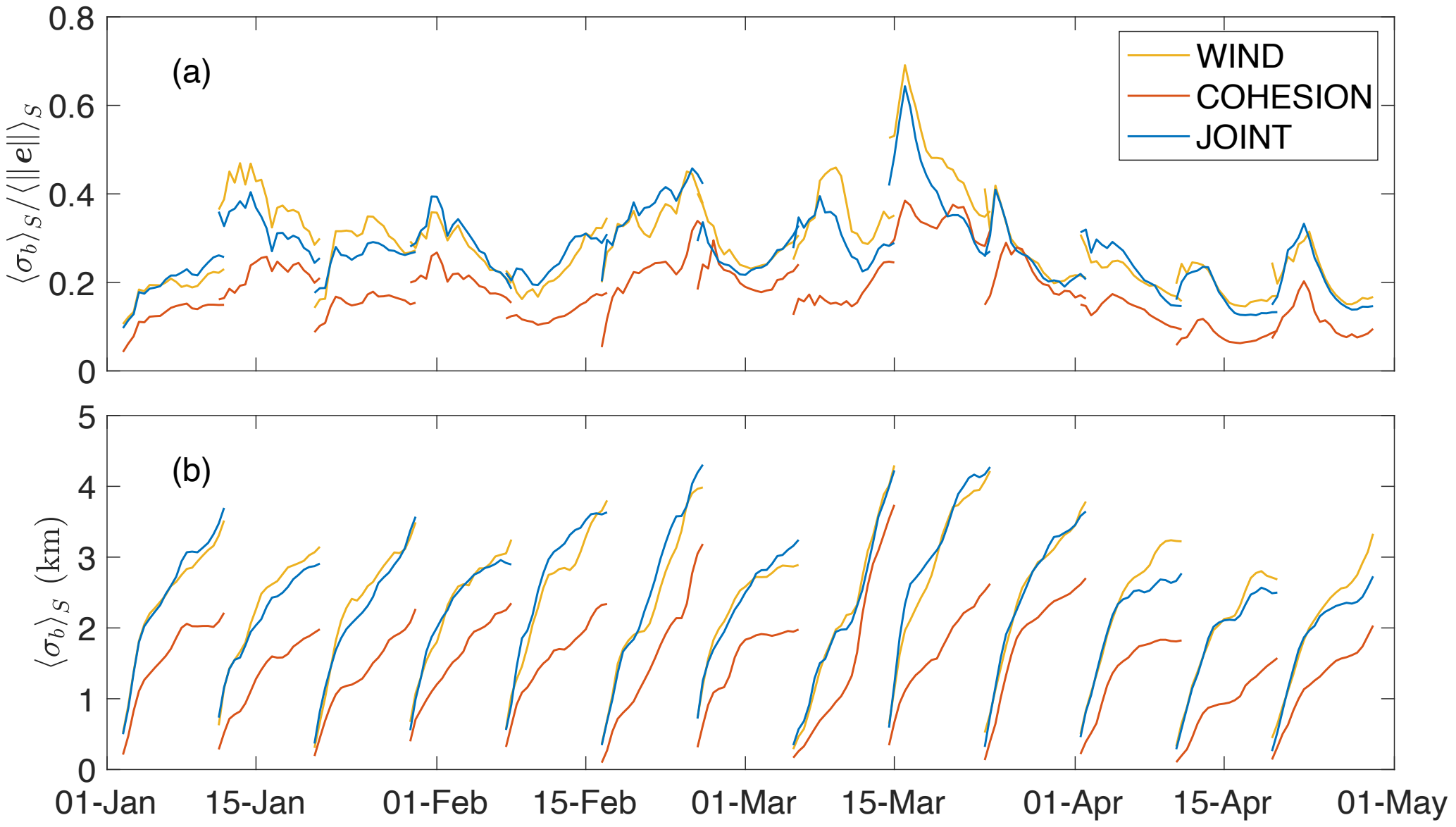

**Figure 5**. (a) Spatial average of spread over RMSE ratio against time, $\langle \|\sigma_b(t)\| \rangle_S / \langle \|\boldsymbol{e}(t)\| \rangle_S$. (b) Spatially averaged ensemble spread.

### *4.2. Assessment of model sensitivity*

The sensitivity of forecast is assessed using the statistics of ensemble trajectories. We look at the uncertainties in the ensemble forecasts (WIND, COHESION and JOINT) using the trajectories of the second set of the virtual drifters mentioned in section 3.3, which are seeded evenly (50 km distance from each other) over the whole Arctic domain at the beginning of each forecast. At any time and location, we calculate the area and the aspect ratio of the search ellipse by Eq. (4).

The temporal evolution of the ellipse area indicates the diffusion properties of the virtual drifters that are initialized at the same location. Figure 6 presents the spatially averaged ellipse area over time, $\langle A(t) \rangle_S$. The mean and standard deviation of $\langle A(t) \rangle_S$ are indicated as the solid lines and the error bars respectively; colors show different experiments. In this analysis, data exceeding three times the median absolute drift are removed as outliers. The amount of the outliers (61) is low and only 2.2% of the whole data (2796). $\langle A(t) \rangle_S$ is further averaged over the 13 forecast periods as $\langle \langle A(t) \rangle_S \rangle_T$, and displayed in the inset. Figure 6 shows that the ellipse area increases monotonically with time as expected. The estimates from WIND and JOINT are close to each other and are on average 4 times larger than in COHESION. Besides, the lower typical areas (one standard deviation below average) from WIND and JOINT are generally larger than the higher areas (one standard deviation above) from COHESION. This signifies that the dominant forecast uncertainties are due to the perturbation of the wind in most locations, which will be further confirmed below. There are however exceptions to this rule as the second week of March the COHESION ellipses grow faster than in previous weeks and the larger ellipses can become larger than the smaller ellipses from both WIND and JOINT.

We calculate the temporal average of $A(t)$ evolved from the same starting positions (i.e., green dots in Figure 1(c)), $\langle A(t)\rangle_T$. Figure 7 shows the spatial distribution of $\langle A(t)\rangle_T$ at forecast horizons 3, 5, and 7 days for each experiment, where $\langle A(t)\rangle_T$ are presented at the starting positions and are scaled by a color bar. The scale of the color bar is capped at 1000 km$^2$ to highlight regional differences in the central Arctic. Throughout the temporal domain, the statistics of the ellipse areas from WIND and JOINT are in agreement, with a growth rate of the ellipse area of about 89 km$^2$/day, while the ellipse area from COHESION is much smaller with a growth rate of about 22 km$^2$/day. In addition, the larger ellipse areas, in all experiments, are found near the ice edge in the Chukchi/Beaufort Sea to the west, as well as the Nansen Basin, Barents Sea and Kara Sea to the east. Strong and narrow currents in those regions may contribute to the increased sea ice diffusion.

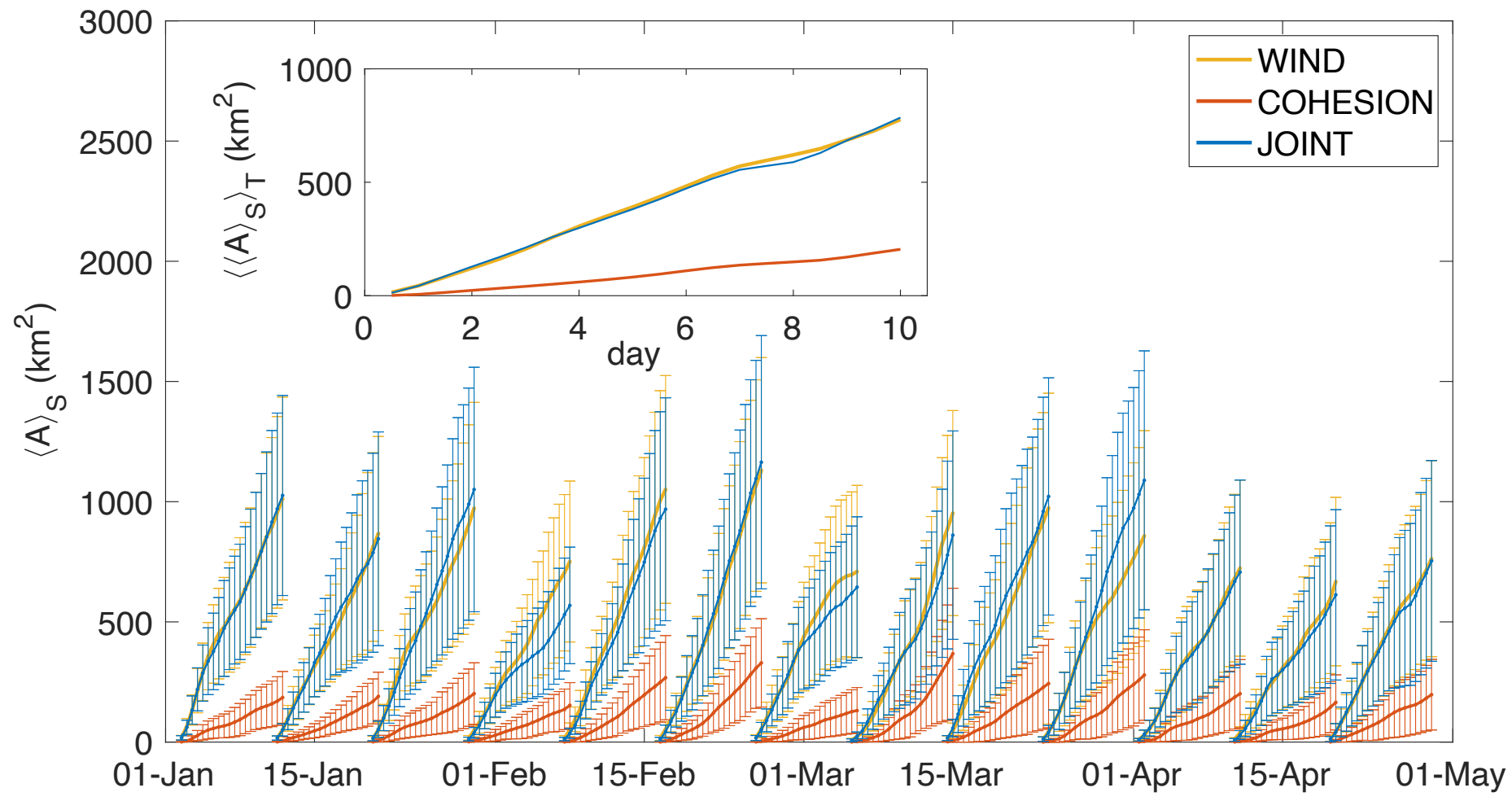

**Figure 6.** Temporal evolution of the spatially averaged area of the anisotropic search ellipses. The solid lines indicate the mean and error bars indicate the standard deviation. Inset: the spatial-temporal domain average of the ellipse areas against lead time.

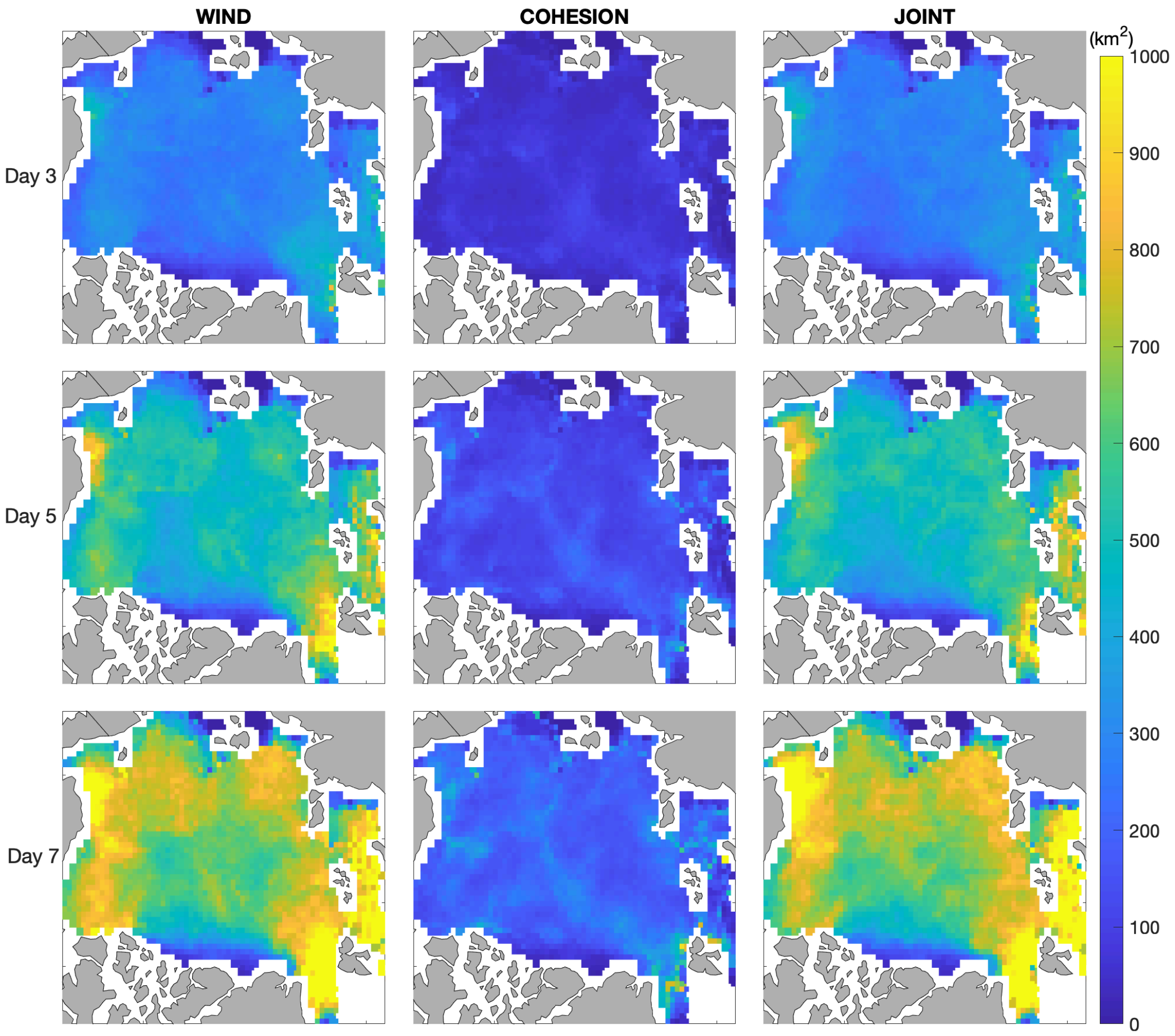

**Figure 7.** Spatial distribution of temporal-averaged ellipse area over all periods at lead days 3, 5, and 7. The color scale is capped at 1000 km² to highlight regional differences, thus very large spreads saturate at the ice edge in the Fram Strait, related to strong local currents.

We further compare the forecast uncertainty generated by the wind perturbation and cohesion perturbation alone. $C(t)$ is defined as the ratio of the intersection area between the wind-ellipse and the cohesion-ellipse over the area of the cohesion-ellipse at time $t$. Hence, $C = 1$ indicates that the wind-ellipse includes completely the cohesion-ellipse, while $C = 0$ indicates the two ellipses are entirely disjoint. Figure 8 shows that $C$ averaged over all virtual drifters is higher than 0.9 throughout the studied period, indicating that the ellipse due to cohesion perturbations is mostly included in wind-driven ellipses. This is the case over nearly the entire Arctic at the exception of an approximately 100 km×100 km region near the coast between north of Greenland and the Canadian Archipelago (not shown): we suspect that this exception could be related to the specific role of sea ice cohesion in areas of very thick ice.

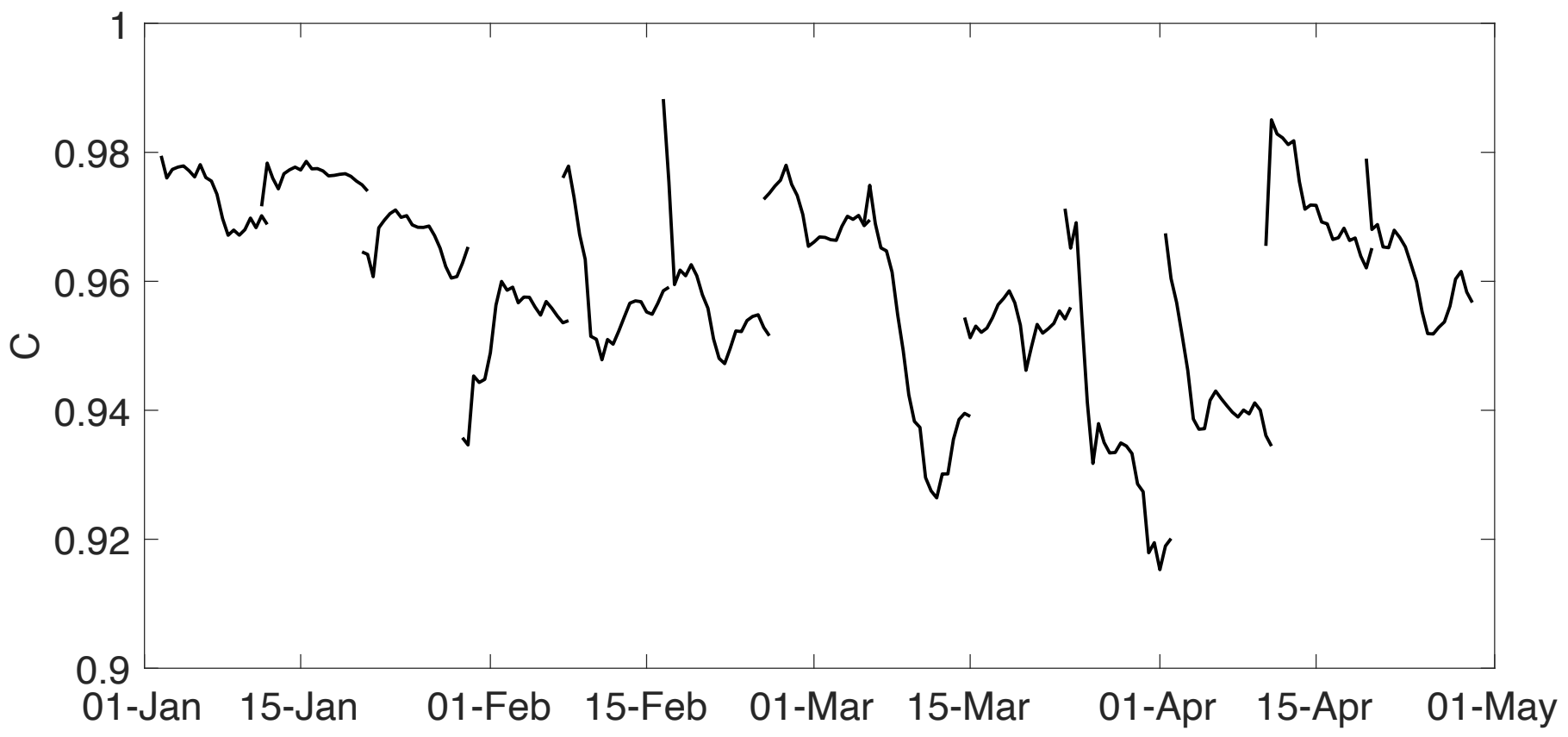

**Figure 8.** Probability of the cohesion-induced forecast uncertainty being included in the wind-induced forecast uncertainty.

Further, we present the spatial average of the anisotropy against time, $\langle R(t)\rangle_S$, in Figure 9. Results show that $\langle R_{COHESION}\rangle_S$ is nearly twice $\langle R_{WIND}\rangle_S$ but $\langle R_{JOINT}\rangle_S$ is only slightly larger than $\langle R_{WIND}\rangle_S$, for all $t$, so that time is dropped from the notation for clarity. This signifies that the impact of the cohesion perturbations on the sea ice shear deformation is predominant and substantial. The same quantity averaged also over all periods $\langle\langle R(t)\rangle_S\rangle_T$, is shown in the inset. The spatial-temporal averaged anisotropy drops quickly after the first two days in all experiments, but $\langle\langle R_{COHESION}\rangle_S\rangle_T$ decreases at a slower rate than $\langle\langle R_{WIND}\rangle_S\rangle_T$ and $\left\langle\langle R_{JOINT}\rangle_S\right\rangle_T$, which is consistent with Figure 11 in Rabatel et al. [17] obtained for their experiment with wind perturbations. The quantitative differences may however stem from the different definitions of the ellipses adopted here (see Section 2) and not only from the differences in the rheology. We speculate that the decreasing behavior of $R$ may be due to that ice floes (tracked by the virtual drifters) tend to move along the same initial fractures from $t = 0$. With diverse ice-broken patterns due to the applied perturbation, significantly different ice fractures could be developed among the ensemble members in the first 2 days. It leads to a more isotropic dispersion (smaller $R$) of ensemble drifters with respect to their barycenter [17].

Figure 10 shows the spatial distribution of the anisotropy averaged over the 13 periods at lead day 7, $\langle R(\text{day} = 7)\rangle_T$, for all the experiments. Similar patterns are observed in other lead times (not shown). The maximum of the color scale is kept at 3 to better visualize the most central part of the Arctic, however the maximum anisotropy does reach 13 near the coasts. In all experiments, the anisotropy is lower in the central Arctic and higher near the Canadian-Greenland and Siberian coasts. The results of WIND and JOINT experiments are also quantitatively in agreement, showing lower anisotropy than that of COHESION. This is due to the dominance of isotropic wind perturbations, which masks the anisotropy caused by the cohesion perturbations. Moreover, the distribution of anisotropy is opposite to the distribution of the ellipse area shown in Figure 7. The dispersions of ensemble drifters in these pinpoints' areas follow both small and high anisotropic shapes. This may be related to the thick ice or to the presence of landfast sea ice in these regions of the Arctic. Both result in fewer ice fractures, opening of leads, and therefore significantly slower sea ice drift and diffusion.

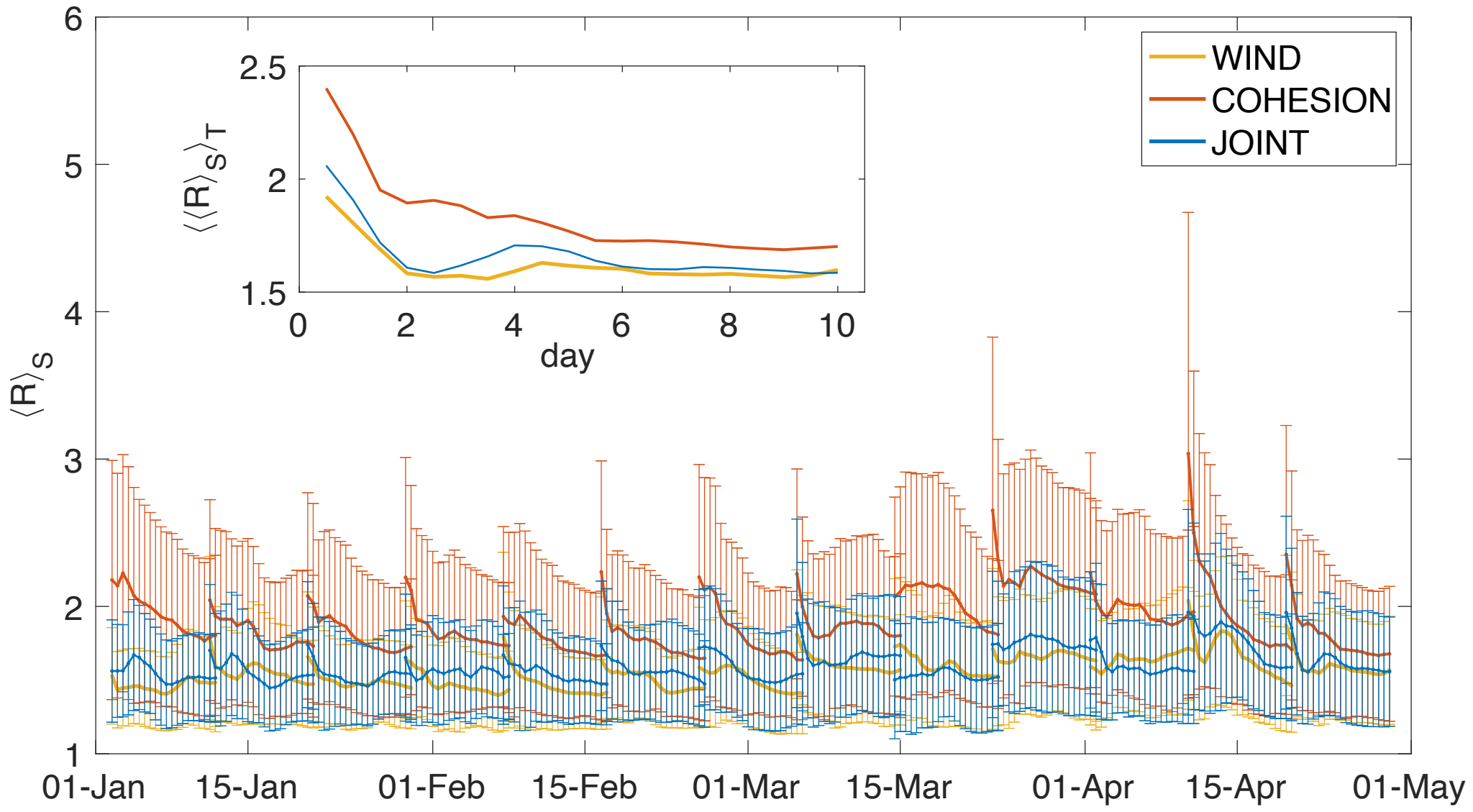

**Figure 9.** Time evolution of the spatial averaged anisotropy of the anisotropic search ellipses. The solid lines indicate the mean and error bars indicate the standard deviation. Inlet: the spatial-temporal domain average of the anisotropy against lead time.

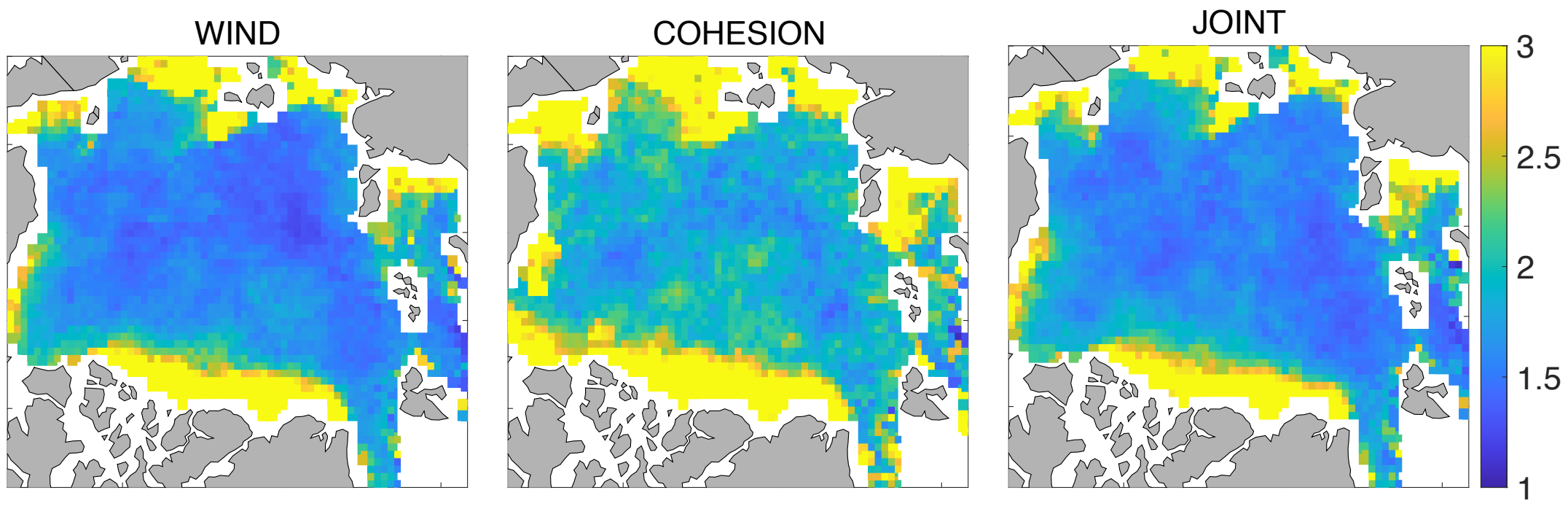

**Figure 10.** Distribution of temporal averaged anisotropy, $\langle R \rangle_T$, over the winter periods at lead day 7.

## 5. Discussion

In this study, we revisited the results from our precursor work [17] using the Lagrangian sea-ice model neXtSIM, this time with the updated Maxwell-Elasto-Brittle rheology of [11]. Compared to the previous EB, the MEB introduces an extra viscous relaxation of stresses mechanism in the damaged sea ice. We evaluated the impact of the assumed main intrinsic (ice cohesion) and extrinsic (surface winds) source of uncertainties on sea ice trajectory forecasts. The MEB rheology has an unprecedented capacity to reproduce the main characteristics of sea ice deformation [30], and therefore, a positive impact in the overall sea ice drift properties and the simulation of sea ice trajectories.

The study domain is the Arctic Ocean, and the time period is from 1 January to 28 April 2008. This 'wintertime' period is characterized by the sea ice cover extending from coast-to-coast across the Arctic Basin. It is thus adequate to demonstrate the role of cohesion on ice damage, an intrinsic parameter of the brittle-like MEB rheology, which plays a major role on the mechanical behavior of sea ice when the latter is packed. We conduct three types of ensemble forecasts. The first one is generated by perturbing the wind, the second by perturbing the ice cohesion, and the third one by perturbing jointly the wind and the cohesion. The sea ice drift is assessed through the analysis of Lagrangian sea ice trajectories simulated by the model, obtained by seeding and tracking the virtual drifters all over the Arctic region.

The results from the three ensemble experiments are first compared to the trajectories of the IABP buoys. The results of WIND generally agree with the work by [17] using an earlier version of neXtSIM that implemented the EB rheology. The forecast errors and the ensemble spreads of dynamic sea ice drift are overall larger with the perturbation of winds than ice cohesion (cf. Figures 4 and 5). The spread among the virtual buoys underestimates the actual errors in all cases, but even more when only cohesion is perturbed. The ice cohesion is known to play an important role for ice drift in a complete ice cover. However, its effect reduces when ice cover is fragmented, e.g., broken due to the wind, in which case the uncertainty of ice drift relates more to the uncertainty on the winds. This is exemplified by the strong and uniform wind event occurring from 11 to 18 March in the area where the IABP buoys were present. The ensemble spread increases rapidly from 11 to 14 March (cf. Figure 5(b)). The effect is extended to the forecasts from 13 to 22 March: the spreads of WIND and JOINT increase quickly in the first few days. It explains the peak of the ratio of spread over the forecast error in WIND and JOINT on 15 March since the position errors between the simulated drifters and real buoys are small. Bias between the virtual drifters in WIND and the real buoys is larger than that in COHESION, especially during this wind-storm event.

Further, we analyze the dispersion of ensemble drifter trajectories embedded within a fitted ellipse (cf. Eq. (4)). We find that the spatial and temporal characteristics of the diffusion and dispersion in the WIND and COHESION experiment are different, with the WIND ellipses being much larger (Figures 6 and 7), such that the JOINT results mostly resembled the WIND ones. In addition, the ellipses from COHESION are generally contained within the corresponding ellipses from WIND (Figure 8). This confirms that the uncertainties in probabilistic forecasts arise mainly from the wind perturbations. While the above conclusion could quantitatively change under a different choice for the wind and cohesion perturbations amplitude than used here, we speculate that the same qualitative conclusion will still hold under a different choice.

Nevertheless, our results also show that at certain times and locations the cohesion plays an important role and moreover it systematically influences the aspect ratio of the ellipses, that is the degree of anisotropy of the ensemble dispersion. In other words, the individual members tend to disperse along preferred directions (see Figures 9 and 10). We conjecture that this phenomenon is due to ice cracks caused by ice deformations starting from locally low-cohesion regions, where the ice cohesion field is inhomogeneous. Under further internal/external forces, neighboring cracks are connected to form longer linear fractures. As a result, virtual drifters tend to move along the linear features with least resistance from ice internal stress, leading to higher anisotropy in the ensemble as compared to forecasts initialized with a homogeneous cohesion field. Overall, the anisotropy is on average lower in JOINT than in COHESION due to the applied isotropic wind perturbations. It is worth noting that increasing the ensemble size reduces but does not remove the anisotropy. We believe thus that the anisotropy is inherited from the intrinsic mechanical behavior of the sea ice, modulated by the cohesion parameter, and that is characterized by the formation of fractures, geometrical features that are themselves —by nature— anisotropic.

The reduction of anisotropy with time, observed in Figure 9 inset, could be explained as follows. All members of the ensemble initially share identical sea ice properties, including the sea ice damage. During the first few steps of the forecast run, all the drifters starting from the same position in an ensemble tend to drift in the same direction, along the preexisting linear kinematic features, although at different velocities. Thus, the ellipse generated from the ensemble drifters at early times is highly anisotropic. As the simulation progresses, the ice damage level eventually increases under the effect of further break up events, leading to the formation of new —additional— linear kinematic features in other directions, also different from one ensemble member to another. The drift of the ensemble drifters becomes gradually more isotropic over time.

## 6. Conclusions and perspectives

We demonstrate that using the neXtSIM stand-alone model and applying wind perturbations identical to those in the TOPAZ4 HYCOM forecasting system [23] have a significant general impact on sea ice drift, and lead to a large spread of sea ice trajectories. Wind forcing should be considered as a primary source of uncertainty, not only for the sea ice drift, but also for sea ice concentration, thickness, damage and snow thickness on ice. Although to a lesser extent, uncertainty in ice cohesion also occasionally impacts the forecast skill by enhancing the preferential directions of the local drift as a result of the formation of fractures in sea ice. There are two possible explanations to this result: either the anisotropy brought by perturbations of cohesion did not align with the actual anisotropic features in the field and deteriorated on average the drift forecasts, or the benefits of the perturbations were masked by the variable quality of the model forcing (winds and currents may have been more inaccurate during the passage of Arctic storms, thus annihilating the benefits of higher ensemble spread).

Future efforts should therefore focus on the assimilation of sea ice drift observations, which should reduce position errors for all tracers and help align the simulated with the observed anisotropic features. Further, errors in modeled sea ice thermodynamics will in the longer term affect the same tracers. Fortunately, observations of sea ice concentration are readily available (e.g., ESA SICCI, [31]) and observations of sea ice thickness too, although in the winter months only (e.g., CS2SMOS, [32]). For an efficient multivariate assimilation of these observations, the sources of errors in model thermodynamics should be included in the above framework with perturbations of e.g., surface air temperature, snow precipitation and cloud cover. This will allow the assimilation of observed variables to update the unobserved ones, like in TOPAZ4 [33]. The present framework for ensemble simulations would then provide – in the data assimilation jargon – the ensemble forecast, while the analysis can be obtained by applying a generic ensemble data assimilation package.

We are currently working on implementing the ensemble Kalman filter (EnKF, [24]) in neXtSIM. The EnKF has been successfully used in several geophysical systems to improve the forecasts [34]. Nevertheless, its application to neXtSIM is particularly challenging due to the use of a non-conservative moving (Lagrangian) computational mesh in neXtSIM. This implies that each ensemble member will be defined on a mesh with a different number of nodes at different locations. Aydoğdu et al. [35] developed an innovative modification of the EnKF that handles this situation in 1D and allows for updating the physical variables of the model by using a projection onto a properly defined homogenous, fixed-in-time, reference mesh. Going beyond this, the work by Sampson et al. [36] exploits the approach of [35] but develops a variant whereby the nodes location in the ensemble mesh are also updated, thus maintaining the dynamical consistency between physics and mesh at the core of the Lagrangian meshes like the one used in neXtSIM. We are currently studying the extension of their methodology to 2D and its application to neXtSIM. To this scope, the results in this study suggest that, for the ensemble in the EnKF to maintain enough vital spread among trajectories, it is mandatory to perturb the surface wind forcing and it is strongly advisable to perturb the cohesion parameters.

**Author Contributions:** Conceptualization, all authors; methodology, all authors; software, A.A. and S.C.; validation, all authors; formal analysis, S.C.; investigation, S.C.; resources, A.C., L.B. and P.R.; data curation, A.A. and S.C.; writing—original draft preparation, S.C.; writing—review and editing, all authors; visualization, S.C.; supervision, A.A., A.C., P.R. and L.B.; project administration and funding acquisition, A.C. All authors have read and agreed to the published version of the manuscript.

**Funding:** This research was funded by the DASIM-II grant from ONR (award N00014-18-1-2493), the REDDA project from the Research Council of Norway with grant number 250711. A. C. has been also funded by the UK Natural Environment Research Council award NCEO02004.

**Acknowledgments:** Satellite observations and TOPAZ4 data have been downloaded from CMEMS. ASR data have been downloaded from http://iabp.apl.washington.edu/data.html IABP data have been downloaded from https://climatedataguide.ucar.edu/climate-data/arctic-system-reanalysis-asr The authors acknowledge grants of CPU time and disk storage from the Norwegian Sigma2 supercomputing infrastructure (nn2993k and ns2993k). The technical assistance of Timothy Williams and Einar Ólason with the model has been very helpful.

**Conflicts of Interest:** The authors declare no conflict of interest.